\documentclass[pdflatex,sn-mathphys-num]{sn-jnl}

\usepackage{graphicx}%
\usepackage{multirow}%
\usepackage{amsmath,amssymb,amsfonts}%
\usepackage{amsthm}%
\usepackage{mathrsfs}%
\usepackage[title]{appendix}%
\usepackage{xcolor}%
\usepackage{textcomp}%
\usepackage{manyfoot}%
\usepackage{booktabs}%
\usepackage{subcaption}
\usepackage{algorithm}%
\usepackage{algorithmicx}%
\usepackage{algpseudocode}%
\usepackage{listings}%
\usepackage{tabularx}
\usepackage{hyperref}

\theoremstyle{thmstyleone}%

\theoremstyle{thmstyletwo}%

\theoremstyle{thmstylethree}%

\hypersetup{
    colorlinks=true,
    linkcolor=blue,
    filecolor=magenta,      
    urlcolor=blue,
}

\begin{document}

\title[Toward Quantum Monte Carlo for Sovereign Debt Risk: A proof of concept]{Toward Quantum Monte Carlo for Sovereign Debt Risk: A proof of concept}

\author*[1,2]{\fnm{Jean-Sébastien} \sur{Dessureault}}\email{jean-sebastien.dessureault@uqtr.ca}

\author[2]{\fnm{Alitiana Mijoro} \sur{Barisoa}}\email{Alitiana.Mijoro.Barisoa@uqtr.ca}
\equalcont{These authors contributed equally to this work.}

\affil*[1]{\orgdiv{Mathematics and Computer science}, \orgname{Université du Québec à Trois-Rivières}, \orgaddress{\street{3351 Boul. des Forges}, \city{Trois-Rivières}, \postcode{G8Z 4M3}, \state{Québec}, \country{Canada}}}

\affil*[2]{\orgdiv{LSSI - Laboratory of Signal and System Integration, Electrical and Computer Engineering Department}, \orgname{Université du Québec à Trois-Rivières}, \orgaddress{\street{3351 Boul. des Forges}, \city{Trois-Rivières}, \postcode{G8Z 4M3}, \state{Québec}, \country{Canada}}}

\abstract{We present a proof-of-concept pipeline that applies quantum amplitude estimation to the probability that a sovereign debt-to-GDP ratio exceeds a critical threshold at a given horizon. The pipeline combines a reduced two-factor Gaussian debt model linearized around its mean, quantum state preparation on a $32\times32$ grid, a weighted-adder circuit, an integer comparator, and Maximum Likelihood Amplitude Estimation (MLAE). On a noiseless statevector simulator with 500 shots per Grover power, the circuit estimate ($\hat a \approx 0.1152$) agrees with the circuit's exact target amplitude ($0.1153$), and the observed estimation error is consistent with the amplitude-estimation scaling in the number of oracle calls. The model used here is deliberately simple: it admits a closed-form solution, so the contribution is a demonstration of the encoding and estimation workflow, not a practical advantage over classical methods. We discuss the conditions under which such a pipeline could become relevant and the obstacles that remain.}


\maketitle

\section{Introduction}\label{sec1}

Sovereign debt sustainability assessment is a cornerstone of macroeconomic policy, yet it remains challenged by deep uncertainty in growth trajectories, interest rate dynamics, and geopolitical shocks \cite{imf2023fiscal, BIS2022debt}. Monte Carlo simulation is widely used for this purpose, but its estimation error decreases only as $\mathcal{O}(1/\sqrt{N})$ in the number of samples $N$. For rare-event probabilities, this scaling can impose high computational cost \cite{glasserman2004montecarlo}.

Quantum Amplitude Estimation (QAE), introduced by Brassard et al.\ \cite{brassard2002quantum}, reduces the error scaling to $\mathcal{O}(1/N)$ in the number of calls to the state-preparation oracle, a quadratic improvement in query complexity. This has motivated applications to financial risk analysis \cite{woerner2019quantum, egger2020credit} and option pricing \cite{stamatopoulos2020option}. The theoretical foundation is Grover's search algorithm \cite{grover1996fast}, with later variants that avoid quantum phase estimation, notably Maximum Likelihood Amplitude Estimation (MLAE) \cite{suzuki2020amplitude} and iterative approaches \cite{grinko2021iterative}, which reduce circuit depth and improve robustness to finite sampling \cite{montanaro2020quantum}.

An important caveat accompanies these results. A quadratic query speedup does not by itself imply a practical advantage: the cost of implementing the oracle on quantum hardware, the need for fault tolerance at useful depths, and the strength of optimised classical baselines all matter. Resource analyses indicate that quadratic speedups alone may be insufficient for practical advantage without substantial hardware progress \cite{babbush2021focused, chakrabarti2021quantum}. We keep this caveat in view throughout.

In this paper we assemble and test a complete amplitude-estimation pipeline for a sovereign debt tail probability. Our contributions are modest and specific: (i) an explicit workflow from a two-factor linearized debt model to a quantum circuit with a weighted adder and integer comparator; (ii) an MLAE-based estimator evaluated on a statevector simulator with finite-shot sampling; and (iii) an honest accounting of query costs and of the gap between this toy problem and a setting where a quantum approach would be competitive. The model is simple enough to admit a closed-form answer, which we use as a validation reference.

Section~\ref{sec2} describes the methodology, Section~\ref{sec3} reports results, Section~\ref{sec_disc} discusses limitations and relevance, and Section~\ref{sec4} concludes.

\section{Methodology}\label{sec2}

The pipeline has four stages: classical model reduction, quantum state preparation, quantum arithmetic, and amplitude estimation.

\subsection{Classical Model Reduction and Linearisation}

\paragraph{Debt dynamics.} Debt-to-GDP evolves over a horizon of $T=20$ years from $d_0=0.9$ as 

\begin{align}
    \lambda_t &= \frac{r_t-g_t}{1+g_t}\\
    pb_t &= a + b\, (d_{t-1}-d^{*}) + pb_{\mathrm{policy}}, \label{eq:pb}\\
    d_t &= (1+\lambda_t)\,d_{t-1} - pb_t, \qquad t=1,\dots,T, \label{eq:debt}
\end{align}
where $pb_t>0$ denotes a primary surplus (ratio to GDP). In the full simulation model each of $g_t$, $r_t$ and $pb_t$ carries its own yearly noise term; for the quantum pipeline we reduce the model to two aggregate shocks: the \emph{cumulative} growth shock $G$ and the \emph{cumulative} interest-rate shock $R$, i.e.\ the sums over the horizon of the yearly shocks, assumed independent and Gaussian,

\begin{equation}
   G\sim\mathcal N(0,\sigma_G^2),\quad R\sim\mathcal N(0,\sigma_R^2),\quad \sigma_G=\sigma_g\sqrt T,\ \sigma_R=\sigma_r\sqrt T,
\end{equation}

each spread evenly over the years:
\begin{equation}
    g_t = \bar g + \frac{G}{T},\qquad r_t = \bar r + \frac{R}{T}. \label{eq:gr}
\end{equation}

This reduction drops the yearly noise on the primary balance and the persistence of growth. The event of interest is $d_T>\tau$ with $\tau=0.95$, and the target is $a=\Pr(d_T>\tau)$. Table~\ref{tab:meaning_and_value} lists all symbols and values.

\begin{table}[h!]
    \centering
    \begin{tabular}{c c c}
        \toprule
        \textbf{Symbol} & \textbf{Meaning} & \textbf{Value} \\ \midrule
         $T$ & Horizon & 20 years \\
         $d_0 $ & Initial debt/GDP &  0.90 \\ 
         $d^{*} $ & Target debt ratio in the fiscal rule & 0.6 \\
         
         $a$ & Intercept of the primary-balance (pb) rule & 0.0\\
         $b$ & Feedback of the $pb$ on the debt gap & 0.03 \\
         
         $pb_{\mathrm{policy}}$ & Constant policy effect on $pb$ (\%GDP/year) & 0.0 \\
         $\bar g$, $\bar r$ & Trend growth, mean interest rate & 0.015, 0.025 \\
         $\sigma_g$, $\sigma_r$ & Std.\ of the yearly growth / rate shocks & 0.01, 0.005 \\
         $\mathbb E[G]$, $\mathbb E[R]$ & Means of the cumulative shocks & 0, 0 \\
         $\sigma_G$, $\sigma_R$ & Std.\ of $G$, $R$ ($=\sigma\sqrt T$) & 0.04472, 0.02236 \\
         $\lambda$ (nominal) & $(\bar r-\bar g)/(1+\bar g)$ & 0.009852 \\
         $pb_1$ (nominal) & $b\,(d_0-d^{*})$ & 0.0090 \\
         $d_T$ (nominal, $G=R=0$) & Terminal debt ratio & 0.8978 \\
         $\tau$ & Threshold of the event $d_T>\tau$ & 0.95 \\
         $n$ & Qubits per random factor ($2^n$ grid points) & 5 \\
         $m$ & Qubits of the sum register & 6 \\
         $\{k\}$ & Grover powers & $\{0,1,2,4\}$ \\
         $n_{\mathrm{shots}}$ & Shots per Grover power & 500 \\
         \bottomrule 
    \end{tabular}
    \caption{Model and simulation parameters.}
    \label{tab:meaning_and_value}
\end{table}

\paragraph{Linearisation and sign convention.} Growth and interest rate enter with opposite signs: $\partial d_T/\partial G=\ensuremath{-0.7422}$ and $\partial d_T/\partial R=\ensuremath{+0.7349}$ (central finite differences at $G=R=0$ with a step of $\sigma/2$). To obtain two positive weights for the quantum adder we use the \emph{adverse} variables $u_1=-G$ (low growth is bad) and $u_2=+R$. This is equivalent to reversing the index of the growth register, and it is exact because the Gaussian is symmetric about zero, so the probabilities $p_i$ are unchanged. Each $u_i$ is truncated at $\pm3.0\sigma$ and discretised on $2^{n}=32$ regular points $u_i=\ell_i+s_i x_i$, $x_i\in\{0,\dots,2^n-1\}$, with step $s_1=0.00866$ and $s_2=0.00433$ (Table~\ref{tab:measure_error}). The linearised map is

\begin{equation}
d_T \approx d_{\mathrm{const}} + c_1 x_1 + c_2 x_2,
\label{eq:lin}
\end{equation}
where $c_i=(\partial d_T/\partial u_i)\,s_i>0$ is the increase of $d_T$ per grid step, and $d_{\mathrm{const}}=d_T(0,0)-(c_1+c_2)(2^n-1)/2$ is the intercept at $x_1=x_2=0$, the most favourable corner of the grid (it is not the value at the mean, $d_T(0,0)=0.8978$).

The shocks are discretised on a regular grid with $2^{n_{\text{bits}}}=32$ points per dimension ($n_{\text{bits}}=5$ qubits per variable).


\begin{table}[h!]
    \centering
    \begin{tabularx}{\textwidth }{Xl}
       \toprule
       \textbf{Quantity}  & \textbf{Value}  \\
        \midrule 
         Truncation range of $G$ & $[\ensuremath{-0.13416},\,\ensuremath{+0.13416}]$ ($=\pm 3\sigma_G$) \\
         Truncation range of $R$ & $[\ensuremath{-0.06708},\,\ensuremath{+0.06708}]$ ($=\pm 3\sigma_R$) \\
         Probability mass cut per variable & 0.0027 \\
         Grid step of $G$ / $R$ & 0.00866 / 0.00433 \\
         $\partial d_T/\partial G$, $\partial d_T/\partial R$ (per unit) & \ensuremath{-0.7422} / \ensuremath{+0.7349} \\
         Same, per standard deviation & \ensuremath{-0.03319} / \ensuremath{+0.01643} \\
         $c_1$, $c_2$ (increase of $d_T$ per grid step of $u_1=-G$, $u_2=+R$) & \ensuremath{+0.006424}, \ensuremath{+0.003180} \\
         $d_{\mathrm{const}}$ (intercept at $x_1=x_2=0$) & 0.748922 \\
         Exact $d_T$ at best / worst corner & 0.7624 / 1.0620 \\
         Linear model at best / worst corner & 0.7489 / 1.0467 \\
         $\max|d_T^{\mathrm{lin}}-d_T^{\mathrm{exact}}|$ on the grid & 0.0154 \\
         $\max|d_T^{\mathrm{circuit}}-d_T^{\mathrm{exact}}|$ on the grid & 0.0163 \\
         \bottomrule
    \end{tabularx}
    \caption{Truncation, discretisation and linearisation.}
    \label{tab:measure_error}
\end{table}

\paragraph{Error measures.} We use five reference probabilities. $a_{\mathrm{cont}}$: continuous Gaussians and the exact nonlinear model, by classical Monte Carlo with 2\,000\,000 samples (seed 123), standard error 0.0002. $a_{\mathrm{lin,cont}}$: continuous Gaussians and the linear model, which is Gaussian and has the closed form $a_{\mathrm{lin,cont}}=1-\Phi\bigl((\tau-\mu)/\sigma_d\bigr)$ with $\mu=d_T(0,0)=0.8978$ and $\sigma_d=0.03704$. $a_{\mathrm{trunc}}$: Gaussians truncated at $\pm3\sigma$, exact model. $a_{\mathrm{grid}}$: the $32\times32$ grid with the exact model. $a_{\mathrm{circ}}$: the grid with the linear model and integer arithmetic of the circuit, which is the amplitude the circuit encodes. The modelling error decomposes as
\begin{equation}
E_{\mathrm{model}}=a_{\mathrm{circ}}-a_{\mathrm{cont}}=\underbrace{(a_{\mathrm{trunc}}-a_{\mathrm{cont}})}_{\text{truncation}}+\underbrace{(a_{\mathrm{grid}}-a_{\mathrm{trunc}})}_{\text{discretisation}}+\underbrace{(a_{\mathrm{circ}}-a_{\mathrm{grid}})}_{\text{linearisation + integer}},
\end{equation}
and the estimation error is $E_{\mathrm{QAE}}=\hat a-a_{\mathrm{circ}}$. Table~\ref{tab:measure_error} reports both, consistently with the rest of the paper.

\begin{table}[h!]
    \centering
    \renewcommand{\arraystretch}{1.2}
    \begin{tabularx}{\textwidth }{Xl}
       \toprule
         \textbf{Quantity} & \textbf{Value}  \\
         \midrule
         $a_{\mathrm{cont}}$: continuous Gaussians, exact model & 0.0861 ($\pm$0.0002) \\
         $a_{\mathrm{lin,cont}}$: continuous Gaussians, linear model (closed form) & 0.0793 \\
         $a_{\mathrm{trunc}}$: truncated at $\pm3\sigma$, exact model & 0.0845 \\
         $a_{\mathrm{grid}}$: $32\times32$ grid, exact model & 0.0818 \\
         $a_{\mathrm{circ}}$: grid + linear + integer (circuit target) & 0.0818 \\
         $\hat a$: QAE estimate (single run, seed 1) & 0.0809 \\ \midrule
         Linearisation in the continuum, $a_{\mathrm{cont}}-a_{\mathrm{lin,cont}}$ & \ensuremath{+0.0067} \\
         $E_{\mathrm{model}}=a_{\mathrm{circ}}-a_{\mathrm{cont}}$ & \ensuremath{-0.0043} \\
         \quad truncation, $a_{\mathrm{trunc}}-a_{\mathrm{cont}}$ & \ensuremath{-0.0015} \\
         \quad discretisation, $a_{\mathrm{grid}}-a_{\mathrm{trunc}}$ & \ensuremath{-0.0027} \\
         \quad linearisation + integer on the grid, $a_{\mathrm{circ}}-a_{\mathrm{grid}}$ & \ensuremath{+0.0000} \\
         $E_{\mathrm{QAE}}=\hat a-a_{\mathrm{circ}}$ (single run) & \ensuremath{-0.0009} \\
         $E_{\mathrm{total}}=\hat a-a_{\mathrm{cont}}$ & \ensuremath{-0.0052} \\ \midrule
         RMSE of $\hat a$ vs.\ $a_{\mathrm{circ}}$ (200 repetitions, seeds 0--199) & 0.00105 \\
         Bias / std.\ of $\hat a$ over the repetitions & \ensuremath{-0.00000} / 0.00105 \\
         Oracle calls per run, $n_{\mathrm{shots}}\sum_k(2k+1)$ & 9\,000 ($=500\times18$) \\
         Circuit executions per run & 2\,000 \\
         \bottomrule
    \end{tabularx}
    \caption{Probability $a=\Pr(d_T>\tau)$ with $\tau=0.95$ and its errors ($n=5$, $m=6$, $n_{\mathrm{shots}}=500$).}
    \label{tab:probability_error}
\end{table}

The value $E_{\mathrm{lin,int}}=\ensuremath{+0.0000}$ on the grid is specific to this threshold and grid: it arises because the grid with exact and with linear integer arithmetic flags the same grid points. It must not be read as the linearisation error. The linearisation error in the continuum is $a_{\mathrm{cont}}-a_{\mathrm{lin,cont}}=\ensuremath{+0.0067}$, larger than the estimation error and in the opposite direction to the net modelling error, so that part of the modelling error cancels.

\subsection{Quantum State Preparation}

Each discretised Gaussian is loaded into a register of $n=5$ qubits with the amplitudes $\sqrt{p_i}$ of the truncated, renormalised, discretised Gaussian (the density evaluated at the grid points and normalised), producing

\begin{equation}
|\psi\rangle = \sum_{i=0}^{31}\sum_{j=0}^{31}\sqrt{p_i\,q_j}\,|i\rangle|j\rangle,
\end{equation}
where $p_i,q_j$ are the discretised probabilities of $x_1,x_2$. The loading uses Qiskit's generic \texttt{StatePreparation} (exact amplitude-vector synthesis); we did not use \texttt{NormalDistribution} or the Grover--Rudolph construction. For two registers of five qubits, this loading is inexpensive. We note that for general distributions on many qubits, exact loading can require resources exponential in the number of qubits, which is one of the known bottlenecks of this approach.

\subsection{Quantum Arithmetic: Weighted Addition and Comparison}

Let $\delta$ be the debt resolution of one integer unit, chosen so that the full range of $d_T$ on the grid spans the sum register, $\delta=(c_1+c_2)(2^n-1)/(2^m-1)=0.004726$. The weights of bit $j$ of each register are $w_{1,j}=\mathrm{round}(c_1 2^j/\delta)$ and $w_{2,j}=\mathrm{round}(c_2 2^j/\delta)$, so that the circuit computes
\begin{equation}
S=\sum_{j=0}^{n-1}w_{1,j}\,[x_1]_j+\sum_{j=0}^{n-1}w_{2,j}\,[x_2]_j ,
\end{equation}
with $[x]_j$ the $j$-th bit of $x$, using a \texttt{WeightedAdder}; then $d_T\approx d_{\mathrm{const}}+\delta S$. For $n=5$ the weights are $w_1=[1,3,5,11,22]$ and $w_2=[1,1,3,5,11]$. The maximum is $\max S=\sum_j w_{1,j}+\sum_j w_{2,j}=63$, and the sum register of $m=6$ qubits holds up to $2^{m}-1=63$, so $\max S\le 2^m-1$ and no overflow occurs (the slack is 0; the check is tight because $\delta$ is chosen to use the full range).

An \texttt{IntegerComparator} flags the grid points with $S\ge\tau_{\mathrm{int}}$, where $\tau_{\mathrm{int}}=\lceil(\tau-d_{\mathrm{const}})/\delta\rceil=43$. Because of this rounding the circuit's \emph{effective} threshold is $d_{\mathrm{const}}+\tau_{\mathrm{int}}\delta=0.9521$ instead of $\tau=0.95$; its size depends on $n$ (Table~\ref{tab:sweep}). The comparator sets an objective qubit to $|1\rangle$ on the flagged points, so that the probability of measuring $|1\rangle$ equals $a_{\mathrm{circ}}$.

\subsection{Amplitude Estimation via MLAE}

We estimate $a$ using MLAE \cite{suzuki2020amplitude}, which replaces the quantum phase estimation of canonical QAE by a set of Grover powers and classical maximum-likelihood inference. Let $\mathcal{A}$ denote the state-preparation circuit (loading, adder, comparator). The Grover operator is $Q=-\mathcal{A}S_0\mathcal{A}^\dagger S_f$, where $S_0$ reflects about the all-zero state and $S_f$ flips the sign of states with the objective qubit in $|1\rangle$. For a schedule of powers $\{k\}$ we apply $Q^k\mathcal{A}|0\rangle$ and measure the objective qubit $n_{\mathrm{shots}}=500$ times, recording the number $y_k$ of $|1\rangle$ outcomes. Then

\begin{equation}
\Pr(\text{objective}=1\mid k)=\sin^2\!\bigl((2k+1)\theta\bigr),\qquad \sin^2\theta=a.
\end{equation}
The likelihood is
\begin{equation}
\mathcal{L}(\theta)=\prod_{k}\binom{n_{\text{shots}}}{y_k}\bigl[\sin^2((2k+1)\theta(a))\bigr]^{y_k}\bigl[\cos^2((2k+1)\theta(a))\bigr]^{n_{\text{shots}}-y_k},
\end{equation}
and $\hat a=\sin^2\hat\theta$ with $\hat\theta$ maximising $\mathcal L$ on a grid of 200,001 points of $(0,\pi/2)$. A geometric schedule of powers is what gives MLAE its Heisenberg-type scaling \cite{suzuki2020amplitude}; our schedules $(0,1,2,4,\dots)$ are of this type.

\paragraph{Query cost.} A circuit with Grover power $k$ calls $\mathcal{A}$ (or $\mathcal{A}^\dagger$) $2k+1$ times. The total number of oracle calls for a schedule $\{k_m\}$ with $n_{\text{shots}}$ shots per power is therefore
\begin{equation}
N_{\text{calls}} = n_{\text{shots}}\sum_m (2k_m+1).
\label{eq:calls}
\end{equation}
For the schedule $(0,1,2,4)$ used in the main experiment, $\sum_m(2k_m+1)=1+3+5+9=18$, giving $N_{\text{calls}}=9{,}000$, while the number of circuit executions is $4\times500=2{,}000$. We report oracle calls, not circuit executions, as the cost measure when comparing with classical sampling.

\subsection{Implementation Parameters}

We use $n_{\text{bits}}=5$ qubits per variable ($32\times32=1024$ grid points), $n_{\text{sum}}=6$ sum qubits, Grover powers $(0,1,2,4)$, and 500 shots per power. All simulations use a noiseless statevector simulator with shot-based sampling to emulate finite-measurement noise.

Table \ref{reproducibility} presents the reproducibility of this work:
\begin{table}[h!]
    \centering
    \renewcommand{\arraystretch}{1.3}
    \begin{tabularx}{\textwidth}{lX}
     \toprule
        \textbf{Item} & \textbf{Value} \\
      \midrule
        Python & 3.14.7 \\
       \midrule
       Numpy & 2.5.4 \\
       \midrule
        Qiskit version & 2.5.2 \\
       \midrule
       Aer & not used \\
       \midrule
       RNG & \textit{numpy.random.default\_rng}(PCG64) \\
       \midrule
        State loading & \textit{qiskit.circuit.library.StatePreparation}, amplitudes \(\sqrt{p_i}\) of the truncated, renormalised discretised Gaussian. \\ 
        \midrule
        Arithmetic & \textit{WeightedAdder} + \textit{IntegerComparator} (\textit{qiskit.circuit.library}). \\
        \midrule
        Amplitude estimation & MLAE: A$Q^k, k \in [0,1,2,4]$ grid search over $\theta$ (200,001 points). \\
        \midrule
        Circuit simulation & Exact Statevector (no sampling noise); shots drawn with numpy binominial. \\
        \midrule
        Reference Monte Carlo seed/size & 123 / 2,000,000 \\
        \midrule
        RMSE repetitions/seeds & 200/0..199 \\
       \bottomrule
    \end{tabularx}
    \caption{Reproducibility}
    \label{reproducibility}
\end{table}

\section{Results}\label{sec3}

\begin{figure}[ht]
    \centering
    \begin{subfigure}[b]{0.48\columnwidth}
        \centering
        \includegraphics[width=\linewidth]{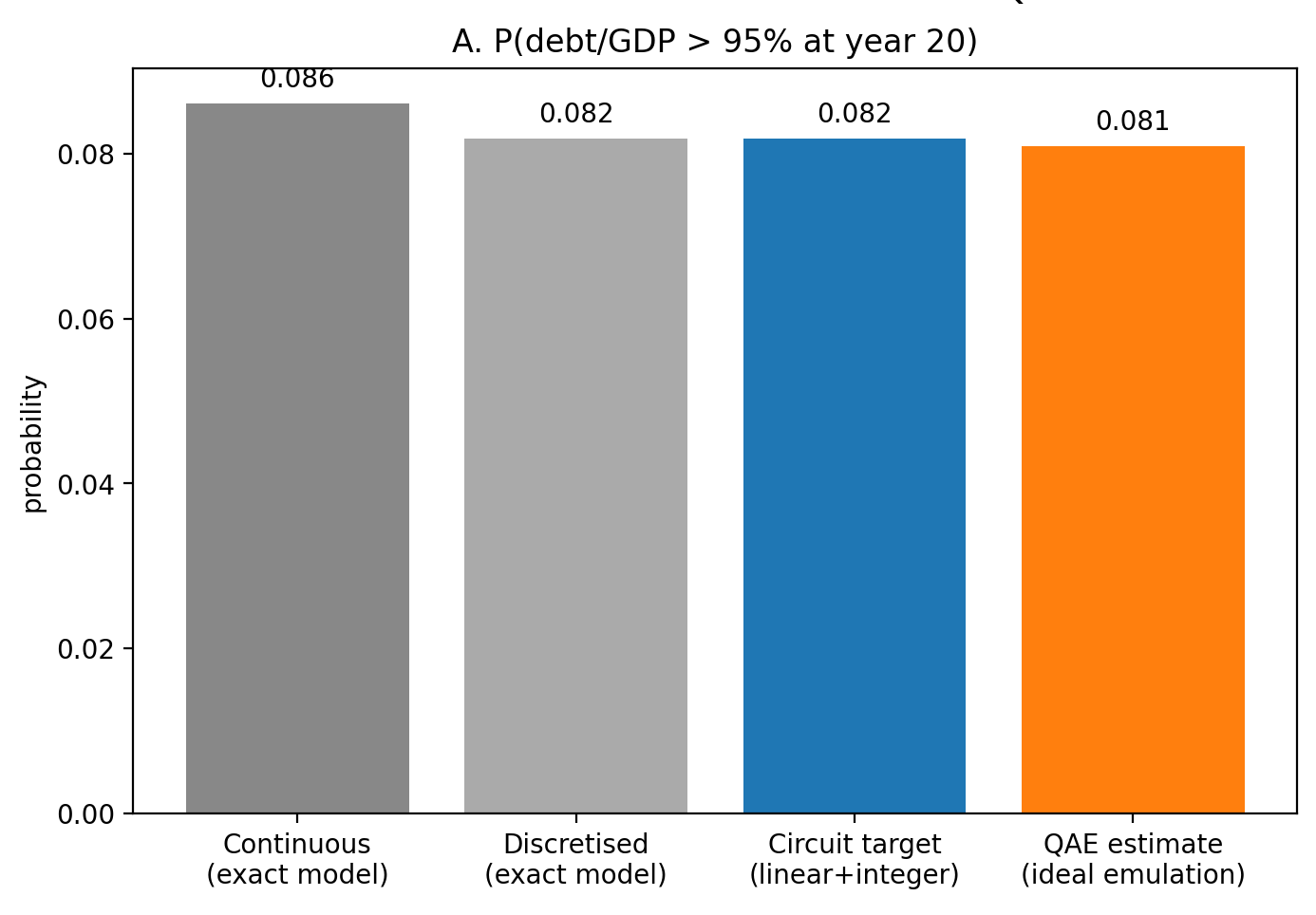}
        \caption{Comparison of probability estimates: continuous exact model, discretised exact model, circuit target (linear + integer arithmetic), and QAE estimate.}
        \label{fig:qmc_a}
    \end{subfigure}
    \hfill
    \begin{subfigure}[b]{0.48\columnwidth}
        \centering
        \includegraphics[width=\linewidth]{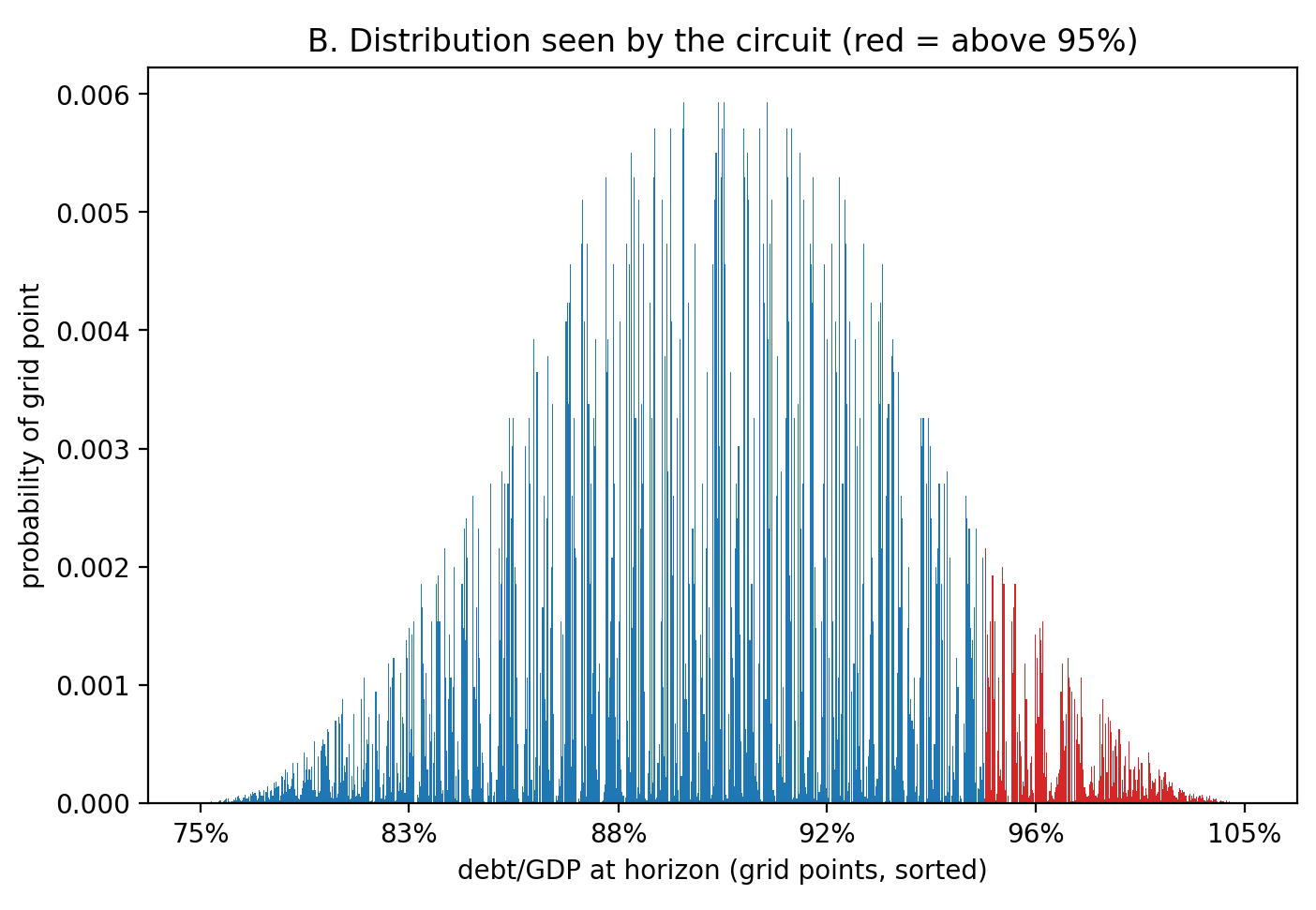}
        \caption{Distribution of $d_T$ encoded in the circuit. Red bars indicate grid points above the 95\% threshold; their total mass equals the target probability $a$.}
        \label{fig:qmc_b}
    \end{subfigure}

    \vspace{0.5em}

    \begin{subfigure}[b]{0.48\columnwidth}
        \centering
        \includegraphics[width=\linewidth]{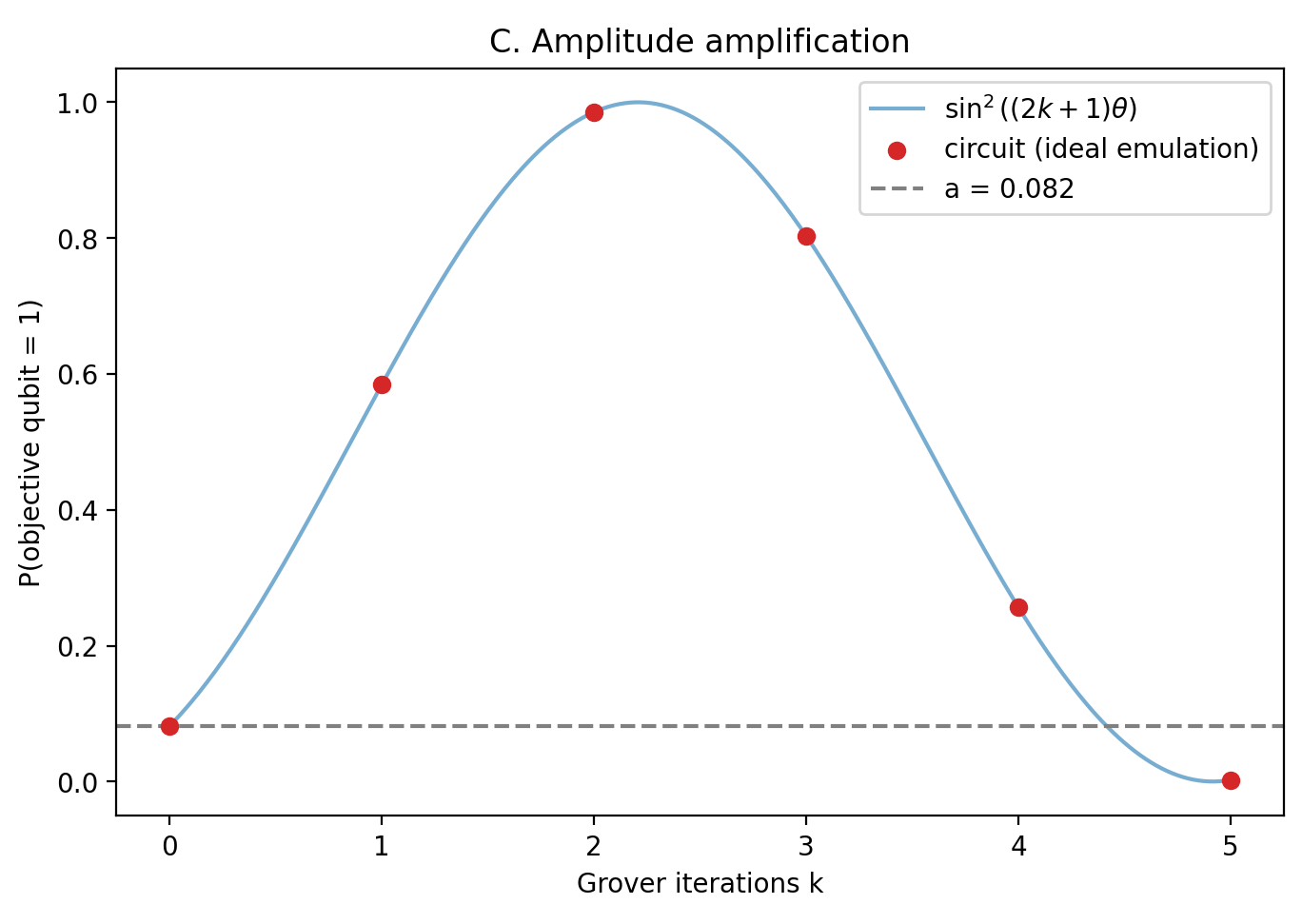}
        \caption{Probability of measuring the objective qubit in state $\lvert1\rangle$ versus Grover power $k$. Red markers: simulation; blue curve: theoretical $\sin^2((2k+1)\theta)$.}
        \label{fig:qmc_c}
    \end{subfigure}
    \hfill
    \begin{subfigure}[b]{0.48\columnwidth}
        \centering
        \includegraphics[width=\linewidth]{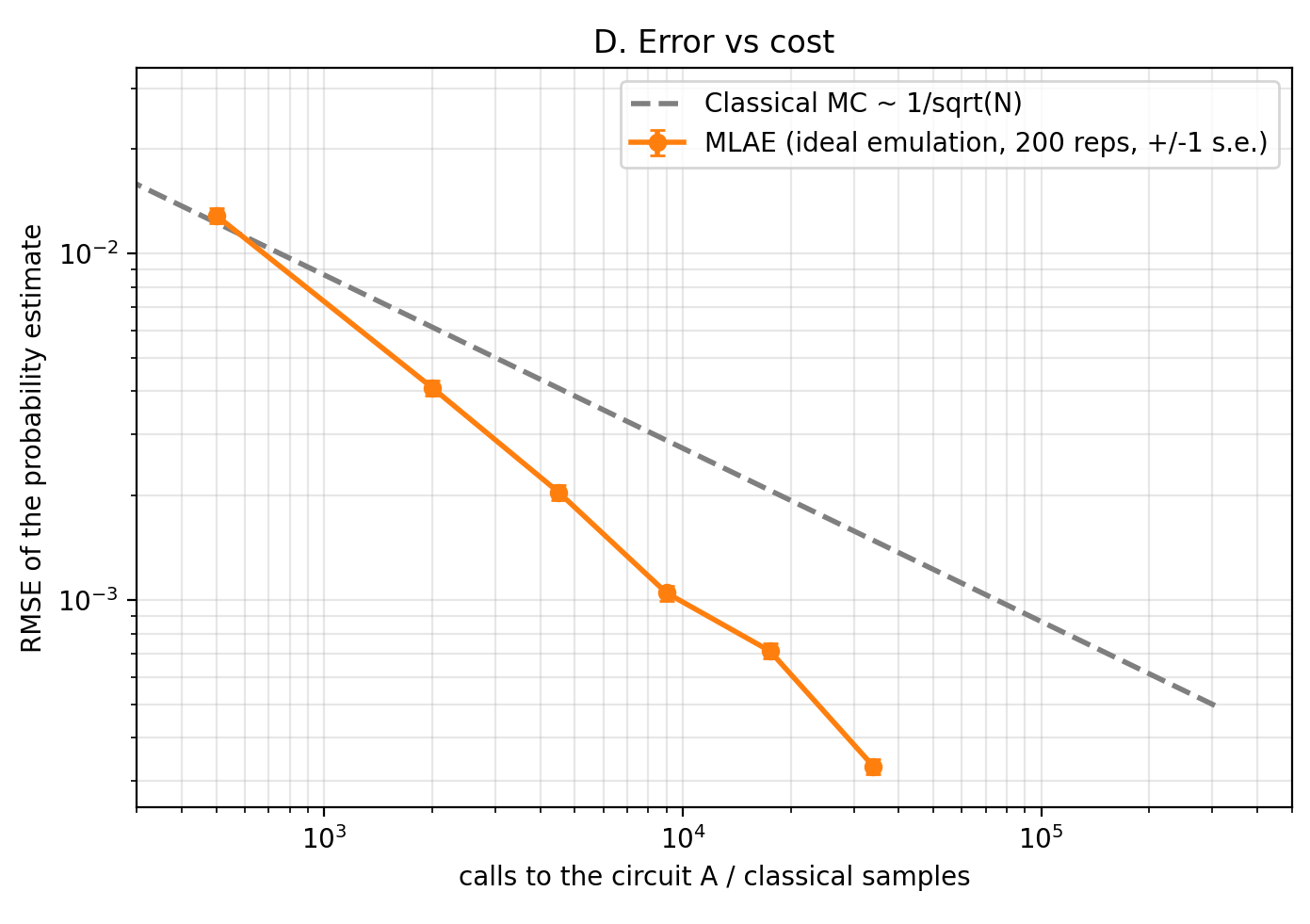}
        \caption{RMSE versus number of oracle calls (log-log). Classical Monte Carlo (grey dashed) scales as $\mathcal{O}(1/\sqrt{N})$; MLAE (orange) is shown for six schedules.}
        \label{fig:qmc_d}
    \end{subfigure}
    \caption{Results for the sovereign debt tail probability. (a) Comparison with reference values. (b) Encoded debt distribution and crisis tail. (c) Grover amplification. (d) Estimation error versus oracle calls.}
    \label{fig:qmc_results}
\end{figure}

Figure~\ref{fig:qmc_results} presents four diagnostics. Figure \ref{fig:qmc_a} Illustrates the baseline precision across successive approximation layers for evaluating the tail event probability (\(P(\text{debt/GDP} > 95\%)\) at year 20). The comparison highlights minimal discretization error between the continuous exact model (\(0.086\)) and the discretized representation (\(0.082\)), with the final MLAE ideal emulation yielding a highly accurate estimate of \(0.081\). \\
Figure \ref{fig:qmc_b} displays the discrete probability distribution mapped across the sorted debt/GDP horizon grid points as encoded by the quantum circuit layout. The targeted regions exceeding the \(95\%\) critical risk threshold are explicitly highlighted in red, visually demonstrating the state preparation phase before amplification. \\
Figure \ref{fig:qmc_c} traces the success probability of the objective qubit as a function of the Grover iteration count \(k\). The ideal emulation data points tightly fit the theoretical \(\sin^2((2k+1)\theta)\) oscillation pattern, validating the coherent amplification of the target state from the initial baseline amplitude of \(a = 0.082\). And the figure \ref{fig:qmc_d} benchmarks the resource efficiency by plotting the RMSE of the probability estimate against the total computational cost (measured via circuit oracle calls or classical sample sizes). The MLAE ideal emulation demonstrates a visibly steeper scaling profile relative to the standard classical Monte Carlo rate (\(\sim 1/\sqrt{N}\)), confirming the anticipated quantum scaling advantage within the tested domain.

\subsection{Agreement with Reference Values}
The single-run MLAE estimate $\hat a=0.0809$ agrees with the circuit target $a_{\mathrm{circ}}=0.0818$ (Fig.~\ref{fig:qmc_a}); over 200 repetitions the estimator has bias \ensuremath{-0.00000} and standard deviation 0.00105 (Table~\ref{tab:probability_error}), so the single-run error $E_{\mathrm{QAE}}=\ensuremath{-0.0009}$ is typical. This confirms that the estimator recovers the amplitude encoded in the circuit. The circuit target in turn differs from the continuous exact model by $E_{\mathrm{model}}=\ensuremath{-0.0043}$, composed of truncation (\ensuremath{-0.0015}), discretisation (\ensuremath{-0.0027}) and linearisation with integer arithmetic on the grid (\ensuremath{+0.0000}); the estimation error is thus 4.1 times smaller than the modelling error. The closed-form probability of the linear Gaussian model is $a_{\mathrm{lin,cont}}=0.0793$, $0.0067$ below $a_{\mathrm{cont}}$.

Table~\ref{tab:sweep} varies the grid size. The modelling error does not decrease monotonically with $n$: the effective threshold fluctuates around $\tau$ because of the integer rounding ($0.9673$, $0.9506$, $0.9521$ for $n=3,4,5$), and the linearisation error in the continuum does not vanish with $n$. Increasing $n$ therefore improves the representation of the Gaussians but not the linearisation, and we do not claim convergence to $a_{\mathrm{cont}}$.

\begin{table}[h!]
\centering
\caption{Dependence on the grid size ($m=n+1$ sum qubits, $n_{\mathrm{shots}}=500$; $\hat a$ from one run, seed 1).}\label{tab:sweep}
\begin{tabular}{cccccccc}
\toprule
$n$ & Grid & $m$ & $\tau_{\mathrm{int}}$ & Effective $\tau$ & $\max S/(2^m-1)$ & $a_{\mathrm{circ}}$ & $a_{\mathrm{circ}}-a_{\mathrm{cont}}$ \\ \midrule
3 & $8\times8$ & 4 & 11 & 0.9673 & 15/15 & 0.0611 & \ensuremath{-0.0250} \\
4 & $16\times16$ & 5 & 21 & 0.9506 & 31/31 & 0.0834 & \ensuremath{-0.0027} \\
5 & $32\times32$ & 6 & 43 & 0.9521 & 63/63 & 0.0818 & \ensuremath{-0.0043} \\
\bottomrule
\end{tabular}
\end{table}


\subsection{Encoded Distribution}
Figure~\ref{fig:qmc_b} shows the probability mass of the linearised debt as computed by the circuit. The total mass of the flagged region (red) equals $a_{\text{circuit}}\approx0.1153$. Note that this is a probability of order $10^{-1}$, i.e.\ not a rare event; the threshold is only about 5.2 percentage points above the nominal terminal ratio $d_T(0,0)=0.8978$.

\subsection{Grover Amplification}
Figure~\ref{fig:qmc_c} shows $\Pr(\text{objective}=1)$ against $k$. The simulated frequencies follow the theoretical $\sin^2((2k+1)\theta)$ curve with $\sin^2\theta=a_{\text{circuit}}$, up to the statistical fluctuation expected for 500 shots. This validates the construction of the Grover operator, with the caveat that the theoretical curve uses $\theta$ obtained from the circuit's own $k=0$ probability.

\subsection{Estimation Error versus Oracle Calls}
Figure~\ref{fig:qmc_d} and Table~\ref{tab:rmse} compare the RMSE of the MLAE estimate with the classical Monte Carlo error $\sqrt{a(1-a)/N}$, as a function of the number of oracle calls $N_{\text{calls}}$ from Eq.~\eqref{eq:calls}, for six schedules, with 200 independent repetitions per schedule (seeds 0--199) and 500 shots per power. The standard error of each RMSE is approximated by $\mathrm{RMSE} / \sqrt{2(n_{\mathrm{rep}}-1)}$


\begin{table}[h!]
\centering
\caption{RMSE versus oracle calls ($a_{\mathrm{circ}}=0.0818$).}\label{tab:rmse}
\begin{tabular}{lrcc}
\toprule
Schedule & $N_{\mathrm{calls}}$ & MLAE RMSE ($\pm$ s.e.) & Classical RMSE at $N=N_{\mathrm{calls}}$ \\ \midrule
$(0)$ & 500 & 0.01286 ($\pm$0.00064) & 0.01226 \\
$(0,1)$ & 2\,000 & 0.00410 ($\pm$0.00021) & 0.00613 \\
$(0,1,2)$ & 4\,500 & 0.00205 ($\pm$0.00010) & 0.00409 \\
$(0,1,2,4)$ & 9\,000 & 0.00105 ($\pm$0.00005) & 0.00289 \\
$(0,1,2,4,8)$ & 17\,500 & 0.00071 ($\pm$0.00004) & 0.00207 \\
$(0,1,2,4,8,16)$ & 34\,000 & 0.00033 ($\pm$0.00002) & 0.00149 \\
\bottomrule
\end{tabular}
\end{table}

For the schedule $(0)$, MLAE reduces to plain sampling and the two errors coincide (0.01286 versus 0.01226), which is a consistency check. At the main schedule ($N_{\mathrm{calls}}=9\,000$) the MLAE RMSE is 0.00105 against 0.00289 for classical sampling at the same number of calls; at the largest schedule ($N_{\mathrm{calls}}=34\,000$) it is 0.00033 against 0.00149, a factor of 4.5. A least-squares fit on the six points gives a log-log slope of $\ensuremath{-0.85}\pm0.02$ for MLAE, versus $-0.5$ for classical sampling.

We emphasise the scope of this observation. With a fixed number of shots per Grover power and powers up to $k=16$, the experiment covers about 1.8 decades of budget. It cannot, by itself, establish an asymptotic $\mathcal{O}(1/N)$ slope, and the fitted slope ($\ensuremath{-0.85}$) lies between the classical and the quantum rate. We therefore regard Fig.~\ref{fig:qmc_d} as consistent with, but not a demonstration of, the quadratic scaling. The comparison is in oracle calls only: one call of $\mathcal A$ is a circuit, whereas one classical sample costs a few arithmetic operations, and the MLAE circuits are simulated without hardware noise. For reference, the theoretical query complexity of canonical QAE to reach additive error $\varepsilon$ is $\mathcal{O}(1/\varepsilon)$, versus $\mathcal{O}(a(1-a)/\varepsilon^2)$ for classical sampling. For $\varepsilon=0.005$ and $a\approx0.0818$, the classical requirement is about 3\,004 samples, whereas the QAE bound is of order $\pi/(4\varepsilon)\approx157$ up to constants and logarithmic factors; in our data the schedule $(0,1)$ with 2\,000 calls already gives an RMSE of 0.00410.

\section{Discussion and Limitations}\label{sec_disc}

\paragraph{The test problem admits a closed form.} Under independent Gaussian shocks and a linear map, $d_T$ is itself Gaussian, and $\Pr(d_T>\tau)=1-\Phi\bigl((\tau-\mu)/\sigma_d\bigr)$ is available analytically (Table~\ref{tab:probability_error}). Quantum (or classical) Monte Carlo is therefore unnecessary for this linearised problem; the closed form serves here as the validation reference. The contribution of the paper is the encoding and estimation workflow, not a computational gain on this instance.

\paragraph{The event is not rare.} The estimated probability is $0.0818$. The motivating use case, tail events with small probability, is not exercised by the present experiment.

\paragraph{Classical baselines.} We compared against plain Monte Carlo only. Competitive classical alternatives, such as quasi-Monte Carlo, importance sampling, and deterministic quadrature, would likely perform much better on this low-dimensional problem and must be included in any claim of practical advantage.

\paragraph{Oracle cost and hardware.} The quadratic speedup is in the number of oracle calls. The oracle here (loading, adder, comparator) is cheap to simulate but, on hardware, each call has a non-negligible circuit cost, and Grover powers multiply the depth. Whether this circuit is executable on near-term noisy devices has not been tested: our simulations are noiseless and we have not transpiled to a hardware topology. Results on fault-tolerant resource requirements for similar financial problems suggest that useful advantage may require large logical-qubit and gate budgets \cite{babbush2021focused, chakrabarti2021quantum}.

\paragraph{Model limitations.} The first-order linearisation is least accurate in the tails, which is exactly where the target probability is defined; the continuous linearisation error is \ensuremath{+0.0067} in probability, and the linear model deviates from the exact one by up to 0.0154 in $d_T$ on the grid. Its assessment is limited to the present parameters. We assume Gaussian shocks and independence; real debt dynamics involve fat tails, correlation, regime changes and policy feedback. The integer resolution of the adder ($\delta=0.004726$ of GDP) and the rounding of the threshold shift the effective threshold to 0.9521. Extending to nonlinear, multi-period, higher-dimensional models is where an oracle-based approach could in principle become relevant, but it is also where oracle construction becomes the main difficulty.

\paragraph{Estimator schedule.} The Grover schedule $(0,1,2,4)$ was chosen heuristically, and we did not compare MLAE against canonical QAE or iterative QAE on the same problem.

\section{Conclusion and Outlook}\label{sec4}

We presented a proof-of-concept pipeline that encodes a two-factor linearised sovereign debt model into a quantum circuit and estimates a tail probability with MLAE. On a noiseless simulator, the estimator recovers the amplitude encoded in the circuit, and the Grover amplification behaves as theory predicts. We have been explicit that the linearised problem has a closed-form solution, that the event is not rare, that only a naive classical baseline was considered, and that the experiment does not by itself establish an asymptotic quantum advantage.

Natural next steps are: (i) non-Gaussian, nonlinear, multi-period debt dynamics with more risk factors, where closed forms are unavailable; (ii) comparison against variance-reduced classical estimators and against other amplitude-estimation variants; (iii) noise models and hardware experiments; and (iv) fault-tolerant resource estimates.

As a longer-term perspective, an amplitude-estimation routine of this kind could be exposed as a tool to software agents operating within digital-twin or simulation environments, such as the geopolitical sandbox of Dessureault et al.\ \cite{dessureault2026evaluating}. We have not implemented or evaluated such an integration, and we do not claim it as a contribution of this work.

\section*{Code and data availability}
The source code and computational pipeline required to replicate the findings of this study are publicly available on \href{https://github.com/brozovic/quantum\_monte-carlo}{GitHub}. An archival version of the repository, including the exact package versions and random seeds specified in Table \ref{reproducibility}. 

\bibliography{sn-bibliography}

\end{document}